\documentclass[aps,prl,preprint,groupedaddress]{revtex4-1}

\usepackage{amsmath}
\usepackage{amssymb}
\usepackage{graphicx}
\usepackage{mathrsfs}
\usepackage{lineno}
\usepackage{bbm}
\usepackage{bbold}

\begin{document}

\title{Dimensionality reduction via path integration for computing mRNA distributions}

\author{Jaroslav Albert\\
jaroslavalbert81@gmail.com}
\begin{abstract}
Inherent stochasticity in gene expression leads to distributions of mRNA copy numbers in a population of identical cells. These distributions are determined primarily by the multitude of 
states of a gene promoter, each driving transcription at a different rate. In an era where single-cell mRNA copy number data are more and more available, there is an increasing need for fast computations
of mRNA distributions. In this paper, we present a method for computing separate distributions for each species of mRNA molecules, i. e. mRNAs that have been either partially or fully processed post-transcription. The method
involves the integration over all possible realizations of promoter states, which we cast into a set of linear ordinary differential equations of dimension $M\times n_j$, where $M$ is the number of available promoter states and $n_j$ is the mRNA copy number of species $j$ up to which one wishes to compute the probability distribution. This approach is superior to solving the Master equation (ME) directly in two ways: a) the number of coupled differential equations in the ME approach is $M\times\Lambda_1\times\Lambda_2\times ...\times\Lambda_L$, where $\Lambda_j$ is the cutoff for the probability of the $j^{\text{th}}$ species of mRNA; and b) the ME must be solved up to the cutoffs $\Lambda_j$, which are {\it ad hoc} and must be selected {\it a priori}. In our approach, the equation for the probability to observe $n$ mRNAs of any species depends only on the the probability of observing $n-1$ mRNAs of that species, thus yielding a correct probability distribution up to an arbitrary $n$. To demonstrate the validity of our derivations, we compare our results with Gillespie simulations for ten randomly selected system parameters.
\end{abstract}

\maketitle

\section{introduction}

In the last decade, single-cell RNA sequencing techniques have advanced to a point where mRNA distributions can be obtained for thousands of genes with high accuracy \cite{Fiers}.
These type of data offer insights into the stochastic processes that govern gene regulatory networks. For this reason, computational techniques that can interpret these data are in high demand. 
One of the aspects of gene regulation that single-cell RNA data can shed light on is the promoter architectures for individual genes. Knowing the mRNA distribution associated with a gene, it is in principal possible to reverse-engineer the promoter architecture that gives rise to said distribution. One approach to achieving this goal is to compute the mRNA probability distributions (PD) for a large number of promoter architectures and select the one(s) that best fits the data. However, this requires fast methods of computing mRNA PDs.

The two most conventional methods of computing PDs for gene products (predominantly RNA and protein) are: solving the Master equation (ME) \cite{Kampen} and the Gillespie algorithm (GA) \cite{Gillespie}.
What makes these two methods attractive is that they are derived from first principles; in fact, the GA is derived from the ME, which makes them different sides of the same coin. In practice the ME is useful only when solvable analytically or when it can be numerically integrated. New analytic and numerical techniques for
solving the ME are constantly being developed, either by means of improving stochastic simulation algorithms \cite{Gibson, Gillespie2, Cao, Cao2, Cao3}, 
or by solving the ME exactly/approximately \cite{Jahnke, Albert, Albert2, Shahrezaei, Pendar, Bokes, Bokes2, Popovic, Veerman}, or by a mix of the former two \cite{Burrage, Jahnke2, 
Albert3, Albert4, Duso, Alfonsi, Kurasov}. In this paper, we enlarge this list by one.

Our approach is to reduce the ME for the mRNA and the promoter to a separate ME for each mRNA species. This is accomplished thanks to a theorem we have
proven in an earlier paper \cite{Albert2}, which allows one to write the generating function (an alternative representation of the ME) for the mRNA as a modified ME for the promoter. 
In this fashion, the individual probability distributions for any one of the species of mRNA can be computed separately. 

The paper is structured as follows: in section 1 we introduce the physical system under consideration and write down the ME for it.
In section 2, we state the aforementioned theorem without proof and proceed to apply it to the system introduced in section one. We derive the ME
for the individual species of mRNA for arbitrary initial conditions. 
In section 3, we test our method against Gillespie simulations for different promoter architectures and discuss the results, advantages and drawbacks of our method. In the concluding section
we summarize our work.

\begin{figure}
\centering
\includegraphics[trim=0 0 0 1.0cm, height=0.35\textheight]{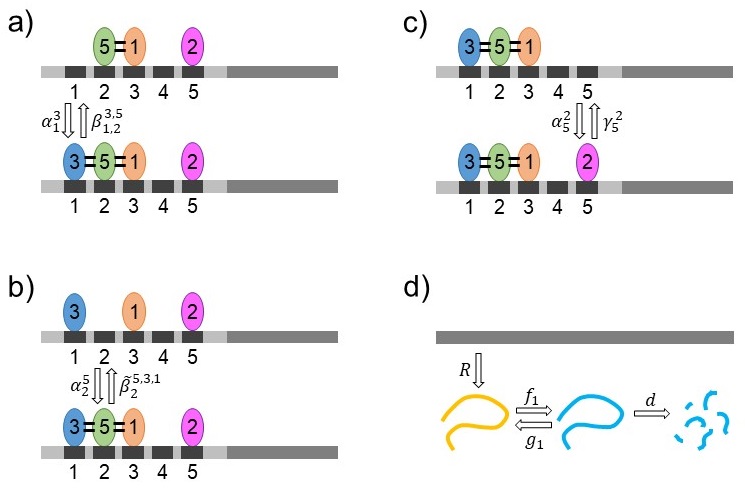}
\caption{A system of a promoter and two species of mRNA. a) Transcription factor 3 binds to and dissociates from
the promoter site 1 at the rate $\alpha^3_1$ and $\beta^{3,5}_{1,2}$, respectively. b) Transcription factor 5 binds to and dissociates from
the promoter site 2 at the rate $\alpha^5_2$ and ${\tilde\beta}^{5,3,1}_2$, respectively. c) Transcription factor 5 binds to and dissociates from
the promoter site 2 at the rate $\alpha^2_5$ and $\gamma^2_5$, respectively. d) Transcription, forward and backward post-transcription process, and mRNA degradation occurring at the rates
$R$, $g_1$, $g_1$ and $d$, respectively.}
\end{figure}

\section{Master equation: direct approach}

The system we wish to describe consists of a gene promoter and mRNA molecules that can be in different post-transcription states. Figure 1 (a-c) shows three possible
promoter states and the processes that cause one state to transform into another. Figure 1 (d) shows the transcription process, the post-transcription processes acting on a newly transcribed mRNA, and
the degradation of a fully processed mRNA. 
If we let $x_i$ be the state of the empty promoter sight $i$, and $y^k_i$ be the state of the promoter sight $i$ occupied by transcription factor (TF) $k$, then the reactions that change the state
of the promoter can be written as
\begin{eqnarray}\label{reactions1}
x_i&\xrightarrow{\makebox[1cm]{$\alpha^k_i$}}&y^k_i\,\,\,\,\,\,\,\,i=1,...,M\nonumber\\
y^k_i&\xrightarrow{\makebox[1cm]{$b^k_i$}}&x_i\,\,\,\,\,\,\,\,i=1,...,M;\,\,\,\,\,\,\,\,k=1,...,N\nonumber\\
\end{eqnarray}
where $\alpha^k_i$ is the TF association rate and 
\begin{equation}\label{b_def}
b_i^k=\gamma_i^kx_{i-1}x_{i+1}+\sum_l\left(\beta^{kl}_{i,i-1}y^l_{i-1}x_{i+1}+\beta^{kl}_{i,i+1}y^l_{i+1}x_{i-1})+\sum_{lp}{\tilde \beta}^{klp}_iy^l_{i-1}y^p_{i+1}\right),
\end{equation}
where $\beta^{kl}_{i,i-1}$ is the dissociation rate of the $k^{\text{th}}$ TF from the promoter site $i$ 
when site $i-1$ is occupied by the $l^{\text{th}}$ TF; $\beta^{kl}_{i,i+1}$ is the dissociation rate of the $k^{\text{th}}$ TF from the promoter site $i$ 
when site $i+1$ is occupied by the $l^{\text{th}}$ TF;
and ${\tilde \beta}_i^{klp}$ is the dissociation rate of the $k^{\text{th}}$ TF from the promoter site $i$ 
when site $i-1$ is occupied by the $l^{\text{th}}$ TF and site $i+1$ is occupied by the $p^{\text{th}}$ TF.
The variables $x_i$ and $y^k_i$ can only take the values 0 and 1. When $x_i=1$,
the $i^{\text{th}}$ promoter site is empty; when $x_i=0$, it is occupied by a TF (any TF). When $y^k_i=1$, the $i^{\text{th}}$ promoter site is occupied by the $k^{\text{th}}$ TF; when
$y^k_i=0$, it is empty. A promoter state is determined by a unique combination of ones and zeros taken by the variables $x_i$ and $y_i^k$, according to the available
promoter sites and the number of TFs trying to bind them. For example, for $M=2$ and $N=2$, a promoter state where TF 1 is bound to promoter site 2, the set of variables
$(x_1,x_2, y_1^1, y_2^1, y_1^2,y_2^2)$
would have the values $(1, 0, 0, 1, 0, 0)$.
For convenience, we define a variable $s$ that labels different promoter states. For example, we could label the state specified by
$(1, 0, 0, 1, 0, 0)$ as $s=1$ and the state specified by $(1, 1, 0, 0, 0, 0)$ as $s=2$. Then, the transition from $s=1$ to $s=2$ would correspond to
a process in which the first TF dissociates from the second promoter site.

The reactions that change the copy numbers of the mRNA species are these:
\begin{eqnarray}\label{reactions2}
&&\emptyset\xrightarrow{\makebox[1cm]{$R(s)$}}m_1\nonumber\\
&&m_j\xrightarrow{\makebox[1cm]{$f_j$}}m_j+1\,\,\,\,\,\,\,\,j=1,...,L-1\nonumber\\
&&m_j\xrightarrow{\makebox[1cm]{$g_j$}}m_j-1\,\,\,\,\,\,\,\,j=2,...,L\nonumber\\
&&m_P\xrightarrow{\makebox[1cm]{$d$}}\emptyset,\nonumber\\
\end{eqnarray}
where $m_1$ is the copy number of the newly transcribed mRNA molecules, $m_j$ ($j>1$) is the copy number of those mRNA molecules that have undergone
$j-1$ post-transcription processes, with $m_L$ being the copy number of the fully processed mRNs; $R(s)$ is the promoter state-dependent transcription rate, $f_j$ is the rate
of conversion from mRNA species $j$ to mRNA species $j+1$, $g_j$ is the rate of the conversion from mRNA species $j$ to mRNA species $j-1$, and $d$ is the degradation rate of mRNA species $L$.
The master equation for the entire system reads
\begin{eqnarray}\label{ME}
\frac{d}{dt}{\bf P}&=&{\bf M}{\bf P}+{\bf R}\left[{\bf P}(m_1-1)-{\bf P}\right]\nonumber\\
&+&\sum_{j=1}^{P-1}f_j\left[(m_j+1){\bf P}(m_j+1,m_{j+1}-1)-m_j{\bf P}\right]\nonumber\\
&+&\sum_{j=2}^Pg_j\left[(m_j+1){\bf P}(m_j+1,m_{j-1}-1)-m_j{\bf P}\right]\nonumber\\
&+&d\left[(m_L+1){\bf P}(m_L+1)-m_L{\bf P}\right].
\end{eqnarray}
We have employed a short hand notation in which ${\bf P}$ is short for ${\bf P}(m_1,m_2,...,m_L,t)$, 
${\bf P}(m_j+1,m_{j+1}-1)$ is short for ${\bf P}(m_1,...,m_j+1,m_{j+1}-1,...,m_L,t)$, etc. The elements of the vector
${\bf P}$, $P_s$, are the probabilities to observe a specific set of copy numbers $(m_1,m_2,...,m_P)$ and the promoter state $s$.
The matrix ${\bf M}$ gives the propensities for transitions between promoter states. Each element of the diagonal matrix ${\bf R}$
gives the transcription rate for a unique promoter state. Since the evolution of the probability of the promoter state does not depend on ${\bf m}$, 
we can sum both sides of Eq. (\ref{ME}) over all $m_j$ to obtain a ME for the promoter:
\begin{equation}
\frac{d}{dt}{\bf \tilde P}={\bf M}{\bf \tilde P},
\end{equation}
where each element of the vector ${\bf \tilde P}$, ${\tilde P}_s$, is the probability to observe the promoter in a state $s$. 

Before we continue, we must establish a connection between the variables $x_i$ and $y^k_i$ and the variable $s$. To do so,
we begin with the ME for the promoter,
\begin{eqnarray}\label{ME_xy}
{\dot P}&=&\sum_i\alpha_i\left[(x_i+1)P(x_i+1,y_i-1)-x_iP\right]\nonumber\\
&+&\sum_ib_i\left[(y_i+1)P(x_i-1,y_i+1)-y_iP\right],
\end{eqnarray}
and define a set
\begin{equation}
{\cal S}=\left\{[x_1,...,x_N],[y_1^1,...,y_N^1],...[y_1^M,...,y_N^M]\right\},
\end{equation}
such that ${\cal S}^s$ represents ${\cal S}$ for particular values of the variables $x_i$ and $y^k_j$.
For example, if $s=1$, we might have
\begin{equation}
{\cal S}^1=\big\{\underset{\underset{\scalebox{1}{$i$}^{\text{th}}\text{\normalsize{site}}}{\uparrow}}{[1,...,0,...,1]},[0,...,0],...,\underset{\underset{\scalebox{1}{$i$}^{\text{th}}\text{\normalsize{site occupied by}} \,\, \scalebox{1}{$k$}^{\text{th}} \text{\normalsize{TF}} }{\uparrow}}{[0,...,1,...,0]},...,[0,...,0]\big\}, 
\end{equation}
which represents a state with the $k^{\text{th}}$ TF bound to the  $i^{\text{th}}$ site. The square brackets
inside ${\cal S}$ are imaginary, serving only as a visual aid; hence, ${\cal S}$ can be thought of as a vector. How we index the promoter states is of no consequence, only that every state has a unique index.
In terms of ${\cal S}$, we can write $x_i={\cal S}^s_i$ and $y^k_i={\cal S}^s_{kM+i}$, where the subscript labels the element of ${\cal S}^s$.
Defining the probability vector as
\begin{eqnarray}
{\bf P}=
\left[\begin{array}{ccccc}
P({\cal S}^1) \\
P({\cal S}^2) \\
\cdot\\
\cdot\\
\cdot 
\end{array}\right],
\end{eqnarray}
the ME (\ref{ME_xy}) can be written in the desired form: 
\begin{equation}\label{ME_xy_to_s}
\frac{dP({\cal S}^s)}{dt}=\sum_{s'}\left\{\sum_{ik}a^k_iC^{ik}_{ss'}(1,-1)+\sum_{ik}b^k_iC^{ik}_{ss'}(-1,1)-\sum_{ik}\left(a^k_i{\cal S}^s_i+b^k_i{\cal S}^s_{kM+i}\right)\right\}P({\cal S}^{s'}),
\end{equation}
where the matrices $C^{ik}_{ss'}(1,-1)$ and $C^{ik}_{ss'}(-1,1)$ are defined as
\[
    C^{ik}_{ss'}(1,-1)= 
\begin{cases}
    1,& \text{if } {\cal S}^{s'}_i={\cal S}^s_i+1, {\cal S}^{s'}_{kM+i}={\cal S}^{s'}_{kM+i}-1\\
    0,              & \text{otherwise}
\end{cases}
\]
\[
    C^{ik}_{ss'}(-1,1)= 
\begin{cases}
    1,& \text{if } {\cal S}^{s'}_i={\cal S}^s_i-1, {\cal S}^{s'}_{kM+i}={\cal S}^{s'}_{kM+i}+1\\
    0,              & \text{otherwise}
\end{cases}
\]
The expression in the curly brackets in Eq. (\ref{ME_xy_to_s}) is the sought after matrix ${\bf M}$. Converting $x_i$ and $y^k_i$ into the new variables ${\cal S}^{s}$ in the dissociation rate, Eq. (\ref{b_def}),
\begin{eqnarray}
b_i^k&=&\gamma_i^k{\cal S}^{s}_{i-1}{\cal S}^{s}_{i+1}\nonumber\\
&+&\sum_l(\beta^{kl}_{i,i-1}{\cal S}^{s}_{lM+i-1}{\cal S}^{s}_{i+1}+\beta^{kl}_{i,i+1}{\cal S}^{s}_{lM+i+1}{\cal S}^s_{i-1})
+\sum_{lp}{\tilde \beta}^{klp}_i{\cal S}^s_{lM+i-1}{\cal S}^s_{pM+i+1},
\end{eqnarray}
completes the switch between the two types of variable.

In principal, Eq. (\ref{ME}) can be solved numerically by imposing upper bounds on all the variables $m_j$, which is not known {\it a priori} but must be
guessed, e. g. by first computing average, ${\bar m}_j$, and the standard deviations, $\sigma_j$, for every $m_j$ (which can be done analytically) and then setting the cutoff to ${\bar m}_j$
plus some multiple
of $\sigma_j$. This {\it ad hoc} way of truncating, however, poses the problem that if the cutoff is too small, the computed probability distribution will be incorrect.
Furthermore, the dimension of the problem, i.e. the number of equations that must be solved, scales as $\Lambda_1\times\Lambda_2\times...\times \Lambda_L\times M$,
where $\Lambda_j$ is the cutoff for the $j^{\text{th}}$ species of mRNA, and M is the dimension of ${\bf M}$ which equals the number of possible promoter states. Given a large enough $L$, and large enough 
$\Lambda_j$s, the task of solving Eq. (\ref{ME}) directly may become computationally unfeasible. 
In the next section, we present a different way of solving Eq. (\ref{ME}), one that reduces the dimension of the problem to $n_j\times M$, 
where $n_j$ is the copy number for the $j^{\text{th}}$ species of mRNA up to which we wish to know the probability distribution of $m_j$.

\section{Master equation: Path integral approach}

\begin{figure}
\centering
\includegraphics[trim=0 0 0 1.0cm, height=0.35\textheight]{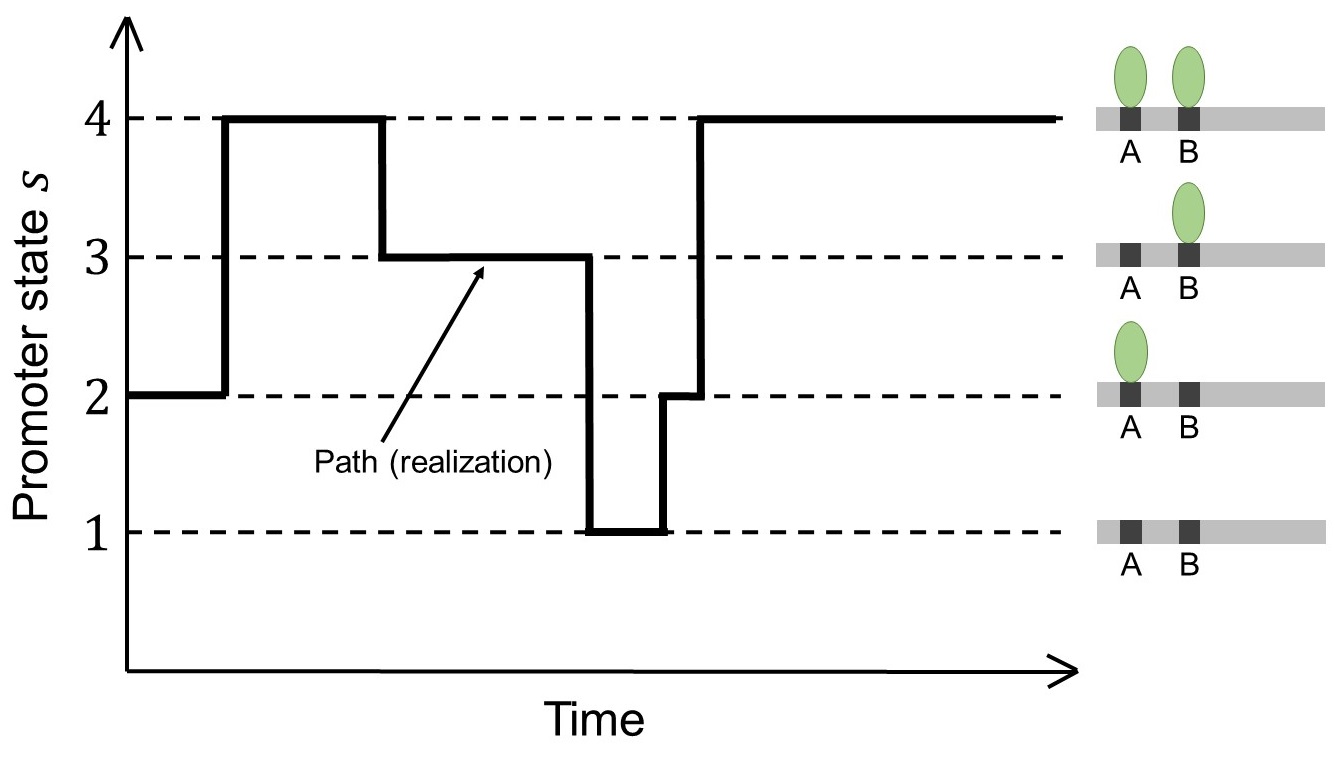}
\caption{An example of a path $s(t)$ for a promoter with two promoter sites acted upon by one TF. The integer values
of $s$ correspond to the following promoter states: $s=1$ - both promoter sites are unoccupied; $s=2$ - only promoter site A is occupied;
$s=3$ - only promoter site B is occupied; $s=4$ - both A and B are occupied.}
\end{figure}

Suppose that we are able to observe the state of the promoter in real time but not the stochastic evolution of the mRNA molecules. We could then write
down a master equation for the variables $m_j$ in which the transcription rate would be a known function of time:
\begin{eqnarray}\label{ME_mRNA}
\frac{d}{dt}P&=&R(t)\left[P(m_1-1)-P\right]\nonumber\\
&+&\sum_{j=1}^{L-1}f_j\left[(m_j+1)P(m_j+1,m_{j+1}-1)-m_jP\right]\nonumber\\
&+&\sum_{j=2}^Lg_j\left[(m_j+1)P(m_j+1,m_{j-1}-1)-m_jP\right]\nonumber\\
&+&d\left[(m_L+1)P(m_L+1)-m_LP\right],
\end{eqnarray}
where $R(t)$ depends on time through the variable $s$: $R(t)=R(s(t))$.
Figure 2 shows an example of what $s(t)$ might look like for a simple promoter with two binding sites and one TF. The evolution of $s$
is sometimes refer to as ``path". 
In what follows, it will be more convenient to work with a generating function (GF), defined as
\begin{equation}
F(\xi_1,...,\xi_L,t)=\sum_{m_1}...\sum_{m_L}\left(\xi_1^{m_1}...\xi_L^{m_L}\right)P({\bf m},t).
\end{equation}
Knowing the GF, one can recover the joined PD from this relation
\begin{equation}
P(m_1,...,m_L,t)=\left[\frac{1}{m_1!...m_L!}\frac{\partial^{m_1}}{\partial \xi_1^{m_1}}...\frac{\partial^{m_L}}{\partial \xi_L^{m_L}}F(\xi_1,...,\xi_L,t)\right]_{{\boldsymbol \xi}=0}.
\end{equation}
Here, we are interested in computing PDs for each variable separately; hence, we will work with a single variable GF, defined as
\begin{equation}
F_j(\xi,t)=\sum_{m_1}...\sum_{m_L}\xi^{m_j}P({\bf m},t),
\end{equation}
from which the single variable PD can be recovered:
\begin{equation}
P(m_j,t)=\left[\frac{1}{m_j!}\frac{\partial^{m_j}}{\partial \xi^{m_j}}F(\xi,t)\right]_{{\boldsymbol \xi}=0}.
\end{equation}
In reference \cite{Albert2}, we have shown that for a system governed by Eq. (\ref{ME_mRNA})
\begin{equation}\label{Ff}
F_j(\xi,t)=G_j(\xi,t)\text{exp}\left[\int_0^t(\xi-1)K_j(t,t')R(t')dt'\right],
\end{equation}
where
\begin{equation}
G_j(\xi,t)=\sum_{n_1}...\sum_{n_L}P({\bf n},0)\prod_{l=1}^M\left[(\xi-1)\sum_iU_{ji}U_{il}^{-1}e^{S_it}+1\right]^{n_l},
\end{equation}
where $P({\bf n},0)$ is the initial joint PD for all variables ${\bf m}$, $S_i$ are the eigenvalues of the matrix
\begin{eqnarray}
{\bf S}=
\left[\begin{array}{cccccccc}
-g_1 & f_2 & 0 & . & . & . & . & 0 \\
g_1 & -(g_2+f_2) & f_3 & 0 & . & . & . & 0  \\
0 & g_2 & -(g_3+f_3) & f_4 & 0 & . & . & 0\\
0 & 0 & g_3& -(g_4+f_4) & f_5 & . & . & 0\\
\cdot & & & & & & &  \\
\cdot & & & & & & &  \\
\cdot & & & & & & &  \\
0 & . & . & . & . & 0 & g_{L-1} & -(f_L+d)
\end{array}\right],
\end{eqnarray}
and ${\bf U}$ is the unit matrix that satisfies $[{\bf U}^{-1}{\bf S}{\bf U}]_{il}=S_i\delta_{il}$.
In the exponent of Eq. (\ref{Ff}), the integration kernel $K_j(t,t')$ is given by
\begin{equation}
K_j(t,t')=\sum_{i=1}^Le^{S_i(t-t')}\left({\bf u}^T_j{\bf U}{\bf B}_i{\bf U}^{-1}{\bf u}_1\right),
\end{equation}
where the elements of the diagonal matrices ${\bf B}_i$
are $[{\bf B}_i]_{lp}=\delta_{li}\delta_{pi}$,
and ${\bf u}_j$ is the $j^{\text{th}}$ unit vector of the bases $[{\bf u}_i]_j=\delta_{ij}$.

Eq. (\ref{Ff}) is valid only for a specific path taken by $s$. 
In order to obtain the PD for the variable $m_j$, regardless of the promoter states, we must multiply Eq. (\ref{Ff}) by the probability of observing a specific path, and then integrate over all possible paths --
a procedure we will refer to as
``integrating (something) over all paths."
This can be accomplished with the help of the following theorem:
\vspace{2mm}
\newline
{\it Theorem 1}: Let ${\bf X}=(X^1,X^2,...,X^V)$ be a set of variables of an arbitrary system,
${\bf X}_i$ be one possible set of values ${\bf X}$ could take, and
\begin{equation}
\frac{d}{dt}P({\bf X},t)-{\cal H}({\cal A}_P,{\bf X},t)=0,
\end{equation}
be the system's ME,
where ${\cal H}$ is some function of ${\bf X}$, $t$ and
${\cal A}_P=(P({\bf X}_1,t), P({\bf X}_2,t),...)$. 
If $P({\bf X},0)$ is the probability to observe ${\bf X}$ at $t=0$, then, for an arbitrary function $W({\bf X}(t'),t,t')$,
integrating
\begin{equation}\label{exp}
\textrm{exp}\left[\int_0^tW({\bf X}(t'),t,t')dt'\right]
\end{equation}
over all paths is given by
\begin{equation}\label{sum}
Q(t)=\sum_{X^1}...\sum_{X^V}Q({\bf X},t')\bigg|_{t'=t},
\end{equation}
where $Q({\bf X},t')$ is the solution of
\begin{equation}\label{Eq_Q}
\frac{dQ({\bf X},t')}{dt'}-{\cal H}({\cal A}_Q,{\bf X},t)=W({\bf X},t,t')Q({\bf X},t')\,\,\,\,\,\,\,\,\,\,\text{for}\,\,\, t'\geq t,
\end{equation}
such that $Q({\bf X},0)=P({\bf X},0)$.  (For proof, see reference \cite{Albert2})
\vspace{2mm}
\newline
In Eq. (\ref{Eq_Q}), $t$ should be considered as a parameter. We will refer to $t'$ as a ``dummy time", since
it is an artefact of the integral in Eq. (\ref{exp}). 
In the present case, ${\bf X}=s$, and ${\cal H}({\cal A}_P,{\bf X},t)={\bf M}{\bf\tilde P}$ and $W({\bf X}(t'),t,t')=(\xi-1)K_j(t,t')$.
Hence, we obtain
\begin{equation}\label{Eq_Q}
\frac{d{\bf Q}(t')}{dt'}=\left[{\bf M}+(\xi-1){\bf R}K_j(t,t')\right]{\bf Q}(t'),
\end{equation}
with the initial conditions ${\bf Q}(0)=G_j(\xi,t){\bf \tilde P}(0)$.
Following the instructions of Eq. (\ref{sum}), we obtain the GF for the variable $m_j$:
\begin{equation}\label{Fcal}
{\cal F}_j(\xi,t)=G_j(\xi,t)Q(t),
\end{equation}
where
\begin{equation}
Q(t)=\left[\sum_{i=1}^M{\bf u}_i\cdot{\bf Q}(t')\right]_{t'=t}.
\end{equation}

Solving Eq. (\ref{Eq_Q}) is not possible; however, we can convert it
into an equation for the PD for $m_j$ by applying the operator $1/(m!)\partial^m/\partial\xi^m$
and then setting $\xi=0$.
The result is this:
\begin{equation}\label{Pm}
\frac{d{\bf P}_m}{dt'}=\left[{\bf M}-{\bf R}K_j(t,t')\right]{\bf P}_m+{\bf R}K_j(t,t'){\bf P}_{m-1},
\end{equation}
where
\begin{equation}
{\bf P}_m=\frac{1}{m!}\frac{\partial^m}{\partial \xi^m}{\bf Q}(t')\bigg|_{\xi=0}.
\end{equation}
Eq. (\ref{Pm}) must be solved for the initial conditions
\begin{equation}\label{InCond}
{\bf P}_m(0)=\left[{\bf \tilde P}(0)\frac{1}{m!}\frac{\partial^m}{\partial \xi^m}G_j(\xi,t)\right]_{\xi=0}.
\end{equation}
To work out Eq. (\ref{InCond}), we can invoke Cauchy's integral formula, which states that
\begin{equation}\label{Cauchy}
\frac{1}{m!}\frac{d^mf(\xi)}{d\xi^m}=\frac{1}{2\pi i}\oint dz\frac{f(z)}{(z-\xi)^{m+1}},
\end{equation}
where $f(z)$ is analytic at the point $\xi$. 
The integral over the complex variable $z$ must enclose $\xi$ but is
otherwise arbitrary. Replacing $f(z)$ in Eq. (\ref{Cauchy}) with $G_j(\xi,t)$, setting $\xi=0$
and performing the integration over a unit circle centered at $z=0$, we obtain
\begin{eqnarray}
\int_0^{2\pi}\frac{d\theta}{2\pi}e^{-mi\theta}G_j(e^{i\theta},t)&=&
\sum_{\bf n}P({\bf n},0)\sum_{q_1=0}^{n_1}...\sum_{q_L=0}^{n_L}\prod_{l=1}^P{n_l \choose q_l}
h_{jl}^{q_l}(1-h_{jl})^{n_l-q_l}\nonumber\\
&\times&\int_0^{2\pi}\frac{d\theta}{2\pi}\text{exp}\left[i\left(\sum_{\mu=1}^Pq_{\mu}-m\right)\theta\right]\nonumber\\
&=&
\sum_{\bf n}P({\bf n},0)\left[\sum_{q_1=0}^{n_1}...\sum_{q_L=0}^{n_L}\prod_{l=1}^P{n_l \choose q_l}
h_{jl}^{q_l}(1-h_{jl})^{n_l-q_l}\delta_{m,{\bar q}}\right],
\end{eqnarray}
where
\begin{equation}\label{Cauchy}
h_{jl}=\sum_iU_{ji}U_{il}^{-1}e^{S_it}
\end{equation}
and ${\bar q}=\sum_{\mu}q_{\mu}$.
Hence, the initial conditions for ${\bf P}_m(t')$ are
\begin{eqnarray}
{\bf P}_m(0)&=&
\sum_{\bf n}P({\bf n},0)\left[\sum_{q_1=0}^{{\tilde m}_1}...\sum_{q_L=0}^{n_P}\prod_{l=1}^P{n_l \choose q_l}
h_{jl}^{q_l}(1-h_{jl})^{n_l-q_l}\delta_{m,{\bar q}}\right]{\bf \tilde P}(0).
\end{eqnarray}

\section{Results and discussion}

\begin{figure}
\centering
\includegraphics[trim=0 0 0 1.0cm, height=0.75\textheight]{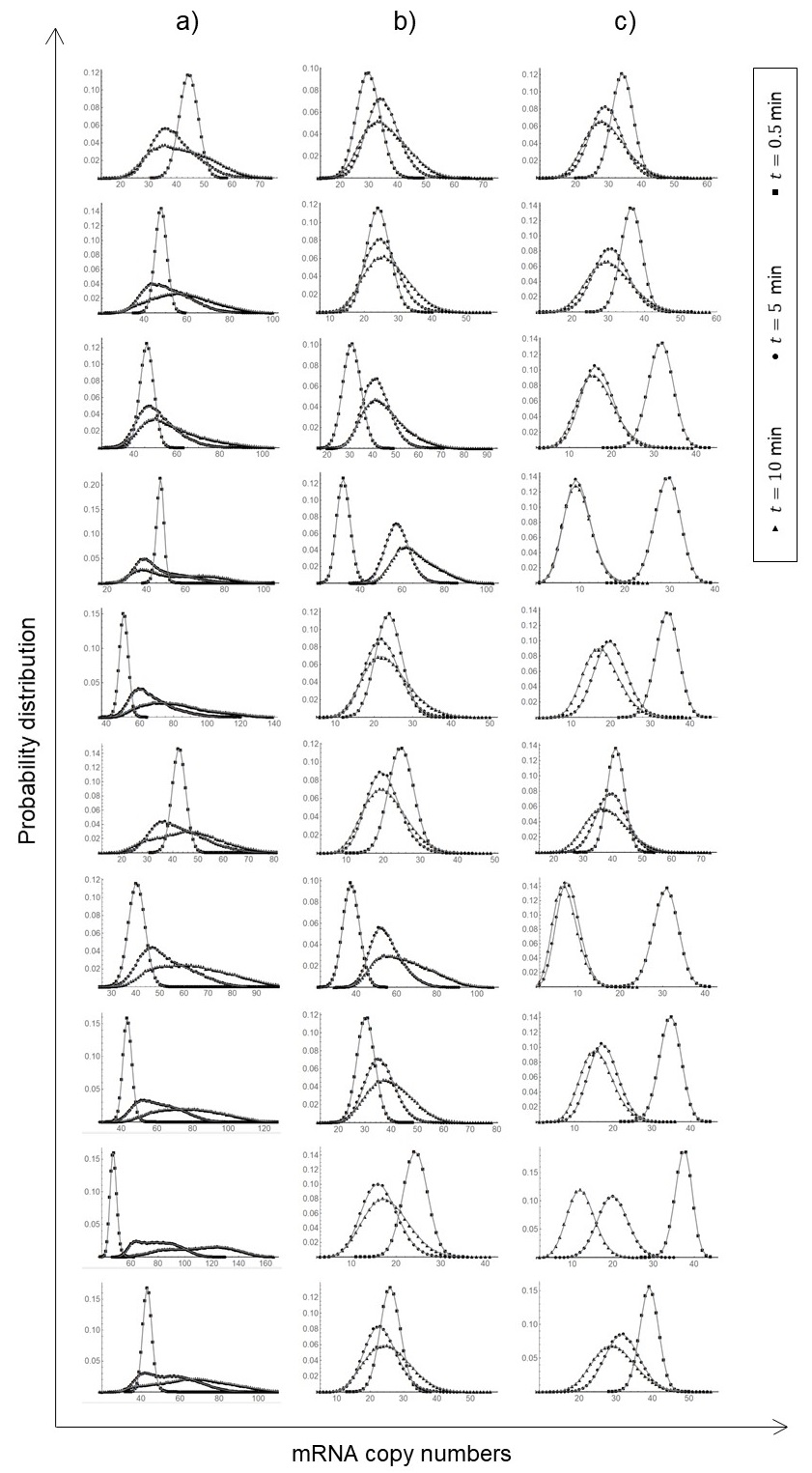}
\caption{Probability distributions for ten randomly selected parameter sets for $M=2$, $N=3$ and $L=3$ at $t=0.5$min (square), $t=5$min (circle)
and $t=10$min (triangle) for a) $m_1$, b) $m_2$, and c) $m_3$. The parameters were sampled from square distributions with the range: $R_{ij}=[1,50]$ min$^{-1}$,
$f_i=[0.05,0.5]$ min$^{-1}$, $g_i=[0.05,0.5]$ min$^{-1}$, $\gamma^k_i=[0.001,0.01]$ min$^{-1}$, $\beta^{kl}_{i,i-1}=\beta^{kl}_{i,i+1}=\gamma^k_i/\kappa_1$ min$^{-1}$,
$\beta^{klp}_i=\gamma^k_i/\kappa_2$ min$^{-1}$, $K_1=(1,2,3,4)$ and $K_2=(1,2,3,4)$. The initial conditions were drawn from square distribution of integers
with the range: $m_1=[0,50]$, $m_2=[0,50]$, $m_3=[0,50]$.}
\end{figure}

In order to test the validity of Eqs. (\ref{Pm}), we generated ten random samples for each of the parameter sets, $\alpha^k_i$, $b^k_i$, $f_i$, $g_i$ and $R_{ij}$
for $M=2$, $N=3$ and $L=3$. For each of the ten cases, we chose initial condition $P({\bf m},0)=\delta_{m_1,{\tilde m}_1}\delta_{m_2,{\tilde m}_2}\delta_{m_3,{\tilde m}_1}$, 
where ${\tilde m}_j$ was randomly selected from square distributions of integers ranging from 0 to 50. Eq. (\ref{Pm}) was solved numerically on Mathematica
using the {\it NDSolve} package for $t=0.5$, $t=5$ and $t=10$. For each parameter set, initial conditions and $t=0.5, 5, 10$, we generated an ensemble of 100k realizations
using the GA, from which we constructed the PD for each variable. The results are presented in Figures 3; the parameter ranges
are given in the figure captions.

The advantage of the method presented herein is that it allows one to decouple the PDs for the mRNA species.
As a result, our method takes us from computationally expensive or infeasible to highly efficient. 
One drawback of this method is that
Eq. (\ref{Pm}) must be integrated over what we termed ``dummy time" from zero to the real time, which must be set beforehand. This means that unlike the solution to the ME,
which, if numerically solvable, gives us a pseudo-continuous solution in time, our method does not. 
To obtain a pseudo-continuous solution in time with our method, one must solve Eq. (\ref{Pm}) for a suitable number of time points and then interpolate the solutions. 
However, in practice, data on probability distributions are usually available only for a few time points; thus, in the context of single-cell mRNA data, our method
is preferable to the ME or the GA.

\section{Conclusion}

We have presented an alternative approach to the ME for a system of an arbitrarily complex promoter and a set of mRNA
species that have either partially or fully undergone the post-transcription processing. The approach consists of obtaining the generating function (GF) for
the mRNAs only as a functional of a particular realization of the promoter state, and then integrating over all possible promoter states.
As a result, we derived an alternative equation for the GF, which we then converted into separate equations for the probability distribution for each species of mRNA for
arbitrary initial conditions.
We have demonstrated the validity of our derivations by comparing the results obtained via our method to those of Gillespie simulations.
This method is highly efficient compared to other methods when the number of mRNA species is greater than one. In practice,
this method lends itself to the reverse-engineering of promoter architectures based on single-cell RNA data.

\end{document}